# Collaborative Domain Blocking:
# Using federated NLP To Detect Malicious Domains

By

MOHAMMAD ISMAIL DAUD

THESIS

Submitted in partial satisfaction of the requirements for the degree of

MASTER OF SCIENCE

in

Computer Science

in the

OFFICE OF GRADUATE STUDIES

of the

UNIVERSITY OF CALIFORNIA

DAVIS

Approved:

Zubair Shafiq, Chair

Setareh Rafatirad

Matthew Bishop

Committee in Charge

2022

**Table of Contents:**

**Abstract:** Current content filtering and blocking methods are susceptible to various circumvention techniques and are relatively slow in dealing with new threats. This is due to these methods using shallow pattern recognition that is based on regular expression rules found in crowdsourced block lists. We propose a novel system that aims to remedy the aforementioned issues by examining deep textual patterns of network-oriented content relating to the domain being interacted with. Moreover, we propose to use federated learning that allows users to take advantage of each other's localized knowledge/experience regarding what should or should not be blocked on a network without compromising privacy. Our experiments show the promise of our proposed approach in real world settings. We also provide data-driven recommendations on how to best implement the proposed system.

## 1.Introduction:

*(1.1 What is ad blocking and why it is needed)* The Web can expose users to a multitude of security and privacy threats. For example, malicious or compromised websites can trick users into downloading malware without having to click on anything via drive-by downloads [26]. As another example, 90% of the top-million websites embed tracking apparatus [15]. To mitigate these security and privacy threats, users employ security and privacy enhancing content filtering tools. For example, 42.7% of users on the internet use ad blockers – making it a popular content blocking technique [2]. An ad blocker browser extension is able to intercept and filter requests to load harmful resources (e.g., iframe elements, JavaScript). They are able to tell a malicious element from a non-malicious element using *filter lists* [5]. These filter lists contain a set of rules in the form of regular expressions to match known security or privacy harming elements on a webpage. As discussed next, filter lists are typically manually curated using crowdsourced feedback from their user base.

*(1.2 Why filterlists are not a cure-all)* While filter lists curation is an impressive community-driven effort that relies on crowdsourced feedback [1], prior literature has shown that the crowdsourced filter list curation process is unreliable [5,12]. There is general consensus that filter list curation is not reliable, especially on less popular and non-English websites [12]. Moreover, Varmarken et al. note that filter lists suffer from both false positives and false negatives. First, these filter lists are known to mistakenly target functional elements that inadvertently break the core functionality of web pages [16]. While such mistakes are expected, it is important for filter list curators to respond and fix these issues in a timely manner. Unfortunately, this is typically not the case. Given the large amount of noisy feedback about website breakage [1], filter list curators have a hard time updating the filter lists in a consistent time frame. Second, filter lists also have a hard time keeping up with new security and privacy threats that require the addition of new rules to filter them [5]. Note that malicious actors are known to employ obfuscation techniques to bypass filter rules [5].

*(1.3 Current ML-Based Alternatives)* Since filter lists are based on crowdsourced, manually generated feedback and suffer from the issues listed above, the research community has employed machine learning to automate filter list generation [11,12] and potentially replace manually curated filter lists altogether [5,6,7,9]. Research has shown that malicious resources tend to have distinct features (e.g. amount of alphabet entropy of malicious element attribute names) that can be leveraged to train accurate machine learning classifiers [5,6,7,9,11,12].

*(1.4 Current ML-Based Alternatives - Augmenting Filter Lists) First,* much of the research looking into blocking malicious content like we are, uses filter lists as a source of ground truth for training their models. However, it is important to note that users can add local rule updates/changes when they find malicious resources not being blocked by the current filter lists they are using. Thus, using only published filter lists as training data forces users to miss out on these local changes and updates that other users within the system are making. Also as discussed earlier, it may take a non-trivial amount of time to get these changes/updates added to the filter list. In order to overcome these issues, we propose a technique known as federated learning. In federated learning, available users are selected to help update a central classifier/model using their own local, private data to generate appropriate updates to it. This central model is then given to all users within the system and is used to classify malicious and non-malicious domains - helping us replace the filter lists based system alluded to earlier. More importantly, since users are proposing direct updates to the model, this allows the system to alleviate the time consuming approval processes associated with updating filter lists and allows us to use list updates/changes that users would not have shared with others to begin with.

(1.5 *Current ML-Based Alternatives - Passive Blocking*) Another common theme seen in prior work is the injection of their proposed models somewhere along the web page rendering pipeline. This type of work uses features that have to be analyzed once the web page loads (e.g number of iframe elements on a page). The elements or pages then get blocked after the page itself has been loaded by the user. Instead of loading in elements of the webpage and then making a judgment based on these elements, we look at complementary network-level resources (i.e whois logs and domain records) that do not require the page to be loaded. So this allows our system to be more passive by nature and more privacy preserving (as we no longer have to analyze the actual contents of the page a user loads). This approach of looking at such information to make an inference on whether its domain is malicious or not also cuts down on needless loading and thereby reduces wasted network resources. This previous statement is especially salient when we end up blocking the page or resource and loading it to begin with was a waste of the network's resources. It also limits the possibility of malicious agents performing a drive-by attack that loads in malware without user consent or action.

(1.6 *Current ML-Based Alternatives - Hand Crafting Features Requires A Lot Of Effort*) Finally, all of the current research tackling this task have one thing in common: the use of expert selected features. The entire feature selection process requires a lot of care and time. These features may also need to get updated as new,complex threats arrive which are no longer detectable by current features sets. A more efficient approach is to use the embeddings of models pre-trained on broadly similar tasks as input features. In our case, we use the BERT model to obtain a compressed and automatically generated representation of the textual input we feed our own models. Research has been done showing the competitive performance of BERT relative to more classical automated text representation techniques [27]. Furthermore, our comparative evaluation with baselines using expert selected features shows comparable performance.

*(1.7 Objective of our system)*

Our objective is to provide a passive, accurate, continually improving and privacy preserving system that can be used to block malicious content at the network level (i.e during the DNS lookup phase). By blocking network requests at this level/stage we can make judgments on content without loading any web page elements. Blocking content at this level also allows us to achieve our secondary objective: adding content blocking capabilities to IoT and mobile devices that seldom have direct access to popular browser based ad blockers.

*(1.8 General ML pipeline)* To train our system to block malicious content during the DNS lookup phase, we gather and process multiple filter lists geared towards IoT and mobile systems from a large database of filter lists (filterlist.com). These filter lists will act as our negative and positive ground truth for what is malicious or non-malicious for our machine learning system. Once we get the domains in a uniform format, we collect the complementary data (i.e whois log and pertinent domain/DNS records) associated with each domain. This machine learning system takes in as input, the whois log of a website and all the associated domains(e.g A,AAAA,CNAME records) of this website. However, before passing in this information to the next step, we process this textual input by passing it through a pre-trained transformer model (i.e BERT) to obtain embeddings we can pass into our two neural network models for classification on whether or not this domain should be blocked or not. One of the neural networks trains on data pertaining to domains written solely by the user for the role of content blocking of domains (i.e. blocking a domain purely for personal and subjective reasons or issues). The other model, which trains on the data pertaining to the gathered domains from the multiple filter lists we collected earlier, actually shares its training gradients with other users within the system through a process known as federated learning. This model aims to block general malicious content like malware, cryptominer, ads,and tracker domains. By using a federated system we allow patterns from local filter list rules to be shared within the system.

*(1.9 Inference pipeline)* The inference pipeline used to predict whether or not a domain is malicious and should be blocked or not, begins with a check to see if a domain is contained within a user's personal content or malicious content filter list. If so, we block the request right there. If not we then pass the whois log and domain/DNS information associated with the requested domain into the malicious content and personal content blocking models. If any one of the models flags the website as undesirable, we block requests to this website. If a mistake is made by the model, the user can update the base filter lists accordingly and allow the models to re-train themselves. In the case of the federated system/model, the model will be allowed to overfitted on these local updates/additions for a couple of rounds before being updated by the globally aggregated model - allowing us to guarantee a system that is at least as good as a purely filter list based system for tail users and majority users alike. Moreover, in the case of the federated model, the distillation of local data patterns on how to block a domain that gets shared globally via gradient sharing and allows all users to take advantage of a global body of knowledge. Finally, using techniques like secure aggregation also ensures the information being shared with the system remains private.

*(1.10 Results)* Through experimentation we are able to view the immediate and promising results of the system. The proposed system was able to achieve comparable performance relative to hand-picked (in terms of features) baseline models and an example system that did not utilize federated learning. This shows a promising road ahead that can lead to further avenues of research and improvement.

## 2. Related Work:

*(2.1 Why do people use ad blockers)* A recent study has shown that approximately 42.7% of users on the internet use ad blockers worldwide [2]. The same study also shows that there has been a steady growth of ad blocker use in the United States over the years [2]. A valid question would be: “why are we seeing this growth”? Factors pushing for this growth and relatively high use of ad blockers are as follows. According to Miroglio et al, people view the presence of ads on webpages negatively due to their association with a bad user experience on the internet [3]. Furthermore, users get the benefit of faster load times of web pages [3], as ads or malicious elements of the pages get blocked before they load themselves onto the web pages. Users also get an added layer of privacy and security [3], as elements that track user activity are blocked and elements that could possibly introduce malicious behavior/agents onto their computer are also blocked.

*(2.2 Issues with filter lists)* At their core, ad blockers implement the policy and/or block lists presented in crowdsourced lists [1]. These rules within these lists can be easily circumvented through techniques of varying complexity [1]. An example of a well-studied technique is the randomization of attributes of elements and URLs served on a page [1]. Since these rules are essentially regular expressions trying to match components, randomizing names and attributes circumvents these regular expressions and thereby circumvents the ad blockers using them. Alrizah, et al (2019) study these circumvention techniques and also expose deeper issues that are created due to the open-source nature of these lists [1]. By looking into the largest and most well known filter list project (i.e EasyList) they were able to pin-point issues that were introducing false positives and negatives into their lists [1]: since anyone can try and contribute to the project, the main editors of the project have a hard time sifting through the low quality reports and additions [1]. It becomes increasingly hard to fix old false positives rules that block unnecessary components or URLs and add rules that inhibit/block the components that are not being correctly blocked(i.e false negatives). Furthermore, websites that have ads and projects like these are essentially natural enemies: when issues of false positives arise website owners seldom collaborate with these lists [1]. Even more worrying, is the introduction of newer circumvention techniques like CNAME cloaking, that mask the actual target domain behind a chain of “dummy” domains [4]. Since these lists are manually curated by volunteers, one cannot expect them to find all the malicious domains across the internet, especially since one domain might actually be hiding behind several others. These circumvention techniques might also further introduce security and privacy issues as they can be associated with improperly set up endpoints and are open to being manipulated or taken over, like some CNAME cloaking instances studied by Dimovaet al (2021) [4]. Essentially, it becomes a lose-lose scenario for internet users with regards to their privacy and security. As stated earlier, the usage of ad blockers is only growing despite these issues being publicly acknowledged. This only points to the biggest issue of them all: there are no other widespread alternatives to ad blockers. This information juxtaposed with the issues expressed in this segment shows that we need a better alternative or reworking of the current system. Thankfully, research is being done into finding such alternatives.

*(2.3 Random Forests To Classify URL)* An appreciable amount of research has already been done on how machine learning can be utilized to block ads on the internet. Lashkari et al (2017) looked into classifying URLs via a server and browser extension system[5]. The URLs would get sent to this server while the browser extension acted on behalf of the client in sending the URLs it encountered to this server [5]. The classifier had an expert selected set of features that were geared to find textual abnormalities within the URL itself (e.g length of the URL,number of tokens, alphabet entropy and many more textual attributes) [5]. The authors of this paper combined these features and fed them into several different classifier algorithms for training, but according to them the most effective algorithm was that of RandomForests(i.e an ensemble of decision trees) [5]. According to the authors, they were able to achieve appreciable performance with this method [5].Though our research also uses a classifier to discriminate between different domains, there are a couple of key differences. The first being that we use word embeddings from a pre-trained BERT model as input: automating the feature creation/selection process and saving human effort/time. The second major difference is that we use a wider set of textual input that is focused on blocking entire domains rather than portions of the website through its URL. We look at the corresponding whois log and DNS records for the base domain instead of just using the URL of the webpage.

*(2.4 kNN To Classify URL)* Bhagavatula et al(2014) also create a very similar classification system with expert-selected features, which is also based on discriminating malicious URLs[6]. Like Lashkari et al, they use textual features and attributes of the URL itself. However, Bhagavatula et al also further the scope of their features by looking into the configuration of the web page that relays information regarding the external or internal URL requests of the page(e.g looking at the proportion of externally requested resources)[6]. The authors of this paper also tried several ML algorithms but ended up stating that the K-Nearest Neighbors algorithms as it yielded the best performance[6]. Again, we take advantage of a more automated feature generation approach that saves time. Furthermore, our whois log feature also allows us to explore the domain in question beyond the confines of the actual domain text. We also get the added benefit of not having to load the entire webpage to extract features as our whois log information is acquired through a third-party API call/command that can be cached.

*(2.5 Decision Tree To Classify URL)* Iqbal et al (2020) propose a system that utilizes application layer data (namely HTML,Javascript and HTTP requests) to create a graph of connections between these resources on a website, in order to classify and block ads [7]. Feature extraction is done on the graph and these features(e.g structural features of the resulting graph) are passed to a decision tree ensemble for classification[7]. The authors of this paper state that the final system was relatively accurate and since the feature extraction component is embedded within the browser's rendering engine component, the system is also efficient [7]. Again, we take the easier and more automated approach in generating our features via our BERT transformation pipeline. Though this system focuses on the overall efficiency and overhead of its implementation by embedding itself within the page rendering pipeline, we still do not require any portion of the web page(instead we look at DNS records and whois logs) to make an inference and thus do not have to waste resources rendering any set of elements that we might end up blocking.

*(2.6 Federated Learning For Malware Detection)* There also has been work on using a federated learning based machine learning system to classify malicious content. Sanket et al. (2021) propose a system to detect malware at a local level using a collection of different machine learning models (e.g CNNs,KNNs) that share their gradients with other users in the system through a federated learning system that collects and aggregates gradients from randomly selected and available users [9]. This "averaged" out gradient is then given to the users who participated in the federated training cycle[9]. Over time, this will allow local, learned patterns to make their way on to other devices - helping the system generalize to a global optimum. They put a heavy emphasis on robustness and anti-poisoning defenses as local users can get corrupted by malware and start corrupting the global federated model by sending out bogus gradients[9]. Namely they use clustering techniques to pick out the possibly malicious gradients[9]. The authors of this paper also stress the energy usage and prediction delay improvements of their formulation - something especially important in IoT driven environments. Our research on the other hand is more focused on network based security and privacy preservation. Though we do try to stop the spread of malware, we do so by blocking potential malware domains rather than focusing on hardware level malware signatures. We also take some inspiration from this work when it comes to creating a robust FL (federated learning) system by checking for gradient updates that are larger than the average values we expect. This stops a couple of users from distorting our system's global optimum it has learnt over time. Our defensive measure also only looks at the actual gradient values of the updates as opposed to looking at extra units of local information(as suggested in Sanket et al.) like the local distribution of features. Thereby further reducing overhead-related inefficiency.

*(2.7 Federated Learning for Malicious Packet Detection)* Bakopoulou et al. (2021) also propose a federated learning system to both stop leakage of private data and ads by classifying(via an SVM model) HTTP packets based on their content [10]. Though one of the core objectives of this research closely aligns with ours (i.e blocking ads), it differs in some fundamental ways. First being that the system takes in application layer packets as input whereas we look at network layer information(i.e domains and whois logs) [10]. The second notable difference is that the system proposed by Bakopoulou et al uses a completely different feature extraction pipeline that looks at HTTP keys, all the words within the packet,and filtered word sets from the packet[10]. They get these words and transform them into a multi-hot encoded vector representing the words shown in the packet. On the other hand, we take the easier/automated approach and pass our textual input into a BERT transformer to capture our inputs in the form of an embedding vector.

*(2.8 Reinforcement Learning for Filter List rule generation)* Hieu et al (2022) took a novel approach and used reinforcement learning to learn a policy that can be used to generate filter lists for websites [11]. The agent ,which is deployed in the environment to learn this underlying policy that makes up filter list creation, is given the ability to detect ads and page usability degradation by adding components representing these ideas into its reward function [11]. According to the authors, this policy successfully generated rules for unseen websites and domains as well and was successful in blocking most ads on popular websites[11]. Such technology could be used to generate filter lists for regions that do not have too many volunteers adding rules to filter list projects. Moreover, this automates the entire labor intensive process of creating filter lists in the first place. There are a couple of limitations though that our research overcomes. First being it still takes a couple of minutes for the agent to generate filter list rules and a person is also required to configure the agent for a given website whereas our approach is more passive, fast and works without intervention due to no configuration being required and more inference pipeline taking a relatively trivial amount of time to generate a prediction. Such technology presented in this work augments the ability of filter lists maintainers rather than outright replacing them. However, it would be interesting and possibly fruitful to combine the work of Hieu et al and the ideas presented in this paper to further improve both systems in tandem via a joint training loop(i.e the RL system provides a constant stream of ground truth that our system can use for training).

*(2.9 Complex Filter List rule generation using page context and behavior)* Alexander et al (2020) propose another system used to generate filter list rules more deterministically(i.e rather than using a learnt policy like above)[12]. The authors of this paper use a hybrid classifier that looks at inter-element relationships and how the element was loaded into the page using a graph structure and also use a convolutional neural network to highlight and flag image-based advertisement elements. The graph structure further adds context to the CNN classified images. Once an element has been classified as an advertisement, a filter list rule is generated by using the graph structure to pinpoint the element's location on the page. According to the authors of this paper, this approach was able to generate novel rules that targeted ads in a way that minimized the breakdown of the web page itself[12]. However, the page must still be technically loaded like it was in Adgraph system references earlier. That is where our approach shines. We can directly cut off possibly malicious requests without ever visiting the webpage by purely analyzing whois log information and domain name information. This makes our approach less obtrusive and more privacy preserving as we no longer have to look into possibly private user generated content on the requested webpage to make a classification. Moreover, we get the added benefit of sharing possibly regional patterns relating to how malicious content is being hosted, with a larger user base through our federated learning system, thus allowing users living in under-represented regions to share information on malicious domains.

*(2.10 The benefits of our system in summary)* As presented above, an appreciable amount of work has been done trying to enhance and improve the current filter list based and dependent system of ad blockers through the introduction of machine learning algorithms and techniques. We build off the strength of these systems and highlight the use of the following mechanisms that stand to further improve the performance of our own machine learning based, content blocking systems. The first element to highlight is the use of a federated learning system that aims to open the possibility of deep pattern sharing amongst users of our system - hopefully allowing everyone to take advantage of each other locally discovered and used filter lists as ground truth. When new types of malicious threats get introduced our ground truth will move in order to block them and our system will follow suit - making the system robust against more global pattern shifts in behavior. Secondly, Our approach is much more passive and does not require complicated representations of web page elements and does not require the loading of resources to make an inference on whether or not a domain is malicious or not.  We achieve this goal of efficiency by only looking at cacheable information that can be acquired through third parties(i.e whois logs and DNS records). Finally, the use of BERT allows us to automatically create a set of features we can feed our models with. This saves on human effort and time that would be spent in researching and crafting the most performant set of features.

**3. Data - Collection & Processing & Patterns:**

*(3.1 Our Main Data Source: Filter Lists)* For the purposes of giving our machine learning system (more specifically the federated model) a source of “ground truth” to discriminate between malicious and non-malicious domains, we use public filter lists hosted on filterlists.com (a large public repository of filter lists). Older related work (e.g. Bhagavatula et al(2014) [6]) and newer work (e.g. Umar et al(2020) [7]), all seem to use filter lists as their ground truth for their proposed systems. This is due to the fact that there are no other widely known alternatives for finding out whether or not a domain is malicious. A possible alternative that we took a cursory look into by testing 4 provably malicious domains if the domain/website trust rating agencies could be reliably used to generate these labels. However, we found that there was a general lack of consensus between the various agencies and extra empirical analysis would be required to separate the more reliable agencies from the bogus ones. Though using filter lists opens the door to the same deficiencies we have previously mentioned, we hope that by using multiple lists the deficiencies of one list will be covered by the content of the others. Furthermore, we only care for the generic patterns (due to the use of our ML oriented approach) used by malicious agents to create domains and our hope is by using multiple lists we will be able to capture said generic patterns.

*(3.2 Data Acquisition and Processing)* As alluded to previously, we utilize filter lists from a large filter list repository,filterlist.com, for training purposes for the federated model portion of our system. However, we did take two different processing procedures for each of our two classes - positive(i.e malicious) and negative (i.e non-malicious).

*(3.3 Positive Class Processing)* For the positive class, we first parsed all the listed projects on filterlist.com and did some basic keyword filtering on title and descriptions of these filter list projects. If either their description or title had a target keyword matching a specific set,say "home" in the IoT word set, we would categorize it accordingly. We had two word sets that we would use to categorize a list as either a mobile or IoT oriented filter list. We did this filtering in order to bring our data in line with our secondary objective: making sure our ML system is oriented towards mobile and IoT network ecosystems. Such ecosystems direly need such a system as most ad blocker software comes in the form downloadable browser extensions, whereas content in mobile and IoT ecosystems can be consumed through non-browser sources like in-built streaming apps. If a list does not contain any target keywords in any of the sets, it is rejected. After this step, we look at the syntax of the filter list and software the list is formatted for. We choose lists that only have DNS-based filter list syntax and associated software that can be used in mobile/IoT environments. Once a list meets both conditions, we check its tag to see what it is set up to block. We only select lists that are oriented to block crypto miners, ads, trackers, malware, and privacy-related leaks/issues. We specifically look at these categories as we will pass this collected and processed data to a federated model that blocks content that is universally seen as malicious - these categories seem to fit this description. After all of this filtering, we parse each list's line/rules and convert them into a uniform format where each domain is written on a new line. We do not format lines that have extra information targeting specific elements on a page or url. We also skipped any allow list rules that allowed domains to bypass lists. We only try to accept/format lines that encode purely domain related information.

**IoT Keywords:**

```
internet of things, internet-of-things, iot, i.o.t, home, pi-hole, pihole, dns,
server, smart, network, router, gateway, protocol
```

| Mobile Keywords: |
|---|
| `ios,android,mobile,phone` |

Table 1: Keywords used to categorize a filter list as either a mobile or IoT filter list(used in positive class processing).

| Software: |
|---|
| `Minerblock - Excluded, AdGuard (free versions), DNS66, AdBlock, AdAway, Pi-hole,FireHOL, Samsung Knox, Privoxy, Diversion, dnsmasq, Blokada, personalDNSfilter, Unbound, BIND, AdGuard Home, pfBlockerNG, Opera’s built-in adblocker, Surge, dnscrypt-proxy, SmartDNS, AdGuard for Android, Vivaldi’s Privacy settings` |
| **Syntax:** |
| `Non-localhost hosts (IPv4), uBlock Origin Static, Domains, Unbound, BIND, Socks5, Hosts (0), Hosts (localhost IPv4), Privoxy action file, Adblocker-syntax domains, Adblocker-syntax domains w/o ABP tag, AdGuard Superadvanced onlys, Adblock Plus, SmartDNS, $important/$empty only, AdGuard, Domains with ABP tags, dnsmasq domains list, Adblock Plus Advanceds, Pi-hole RegEx, Non-localhost hosts (IPv6), DNS servers, Response Policy Zones (RPZ), Domains with wildcards` |

Table 2: Acceptable software and syntax of the filter lists (used in positive class processing).

*(3.4 Positive Class Processing - Downsampling)* After this entire pipeline we still had around 2.5 million domains we could use for the positive class and we had to respect a rate limit for how many whois logs we could access. So, we limited ourselves to taking a maximum of 289 domains per list. This got us to around 14,281 domains for the positive class.

*(3.5 Negative Class Processing)* Due to the dearth of allow lists that matched the criteria we articulated earlier for the positive class, we resorted to taking all of the allow lists we found (regardless if they were IoT or mobile oriented or not). We looked up the syntax “Domains For allow listing” on filterlist.com and listed all the associated lists. We rejected any lists that were part of any acceptable ad program. We do this as there is no broad consensus on whether or not the ads being allow listed in these programs are truly acceptable or not according to potential users[28]. Since this data will be given to a federated model that needs to be trained on data that has broadly agreeable class labels, we skip such lists. In addition to the allow lists collected in the previous manner, we also looked for lists that had “adguard home allow list” as part of their title as we wanted to make sure we got as many IoT/mobile lists as possible and adguard home seemed to be a popular target software on filterlist.com for IoT systems based on a cursory rundown of the data on the site. In the end we got 7 filter lists (allow list) for the negative class and 3 of them were IoT oriented. After getting these lists we parsed each line similarly to the positive class processing pipeline. There was no need for downsampling due to the initial size of our set being relatively small.

*(3.6 Input Processing - Negative and Positive Classes)* Once we have the parsed and cleaned lists we collect the auxiliary information(i.e whois logs and DNS records) associated with each domain that we will actually further process and pass to our federated model. For each domain we query a service to collect its associated whois log (i.e a required log containing information about the domain registrant and other details about the website) and we also collect all of the CNAME and AAAA domains associated with the target domain. We then pass this textual information into a pre-trained BERT transformer model that was trained on tasks requiring it to understand the relationship between words in textual natural language samples[19] ,namely 'bert-base-uncased' of the HuggingFace library, to extract embeddings that will represent our textual information in a format that is usable/readable by our neural network(i.e multi-layer perceptron) model. The whois log is passed line by line to BERT but each line in the log is truncated to the first 512 characters due to the input limit of this model. After we have collected each line's embedding vector we sum each of the vector's columns to get a single output vector and we divide each element in the summed vector by the number of total lines in the log. This essentially averages out the final output vector for the whois log component of our input vector. The domains are split on the "." character and remove any duplicate sub-domains/keywords. We then lexicographically sort all the keywords and join them back together with a space as a delimiter. We then pass this string of sub-domain/keywords into BERT and extract the output embeddings. It is also important to note that we reject an entire domain instance(i.e whois log and DNS record vector concatenated combination of a specific target domain) if we see a blank whois log or if the BERT output of either the DNS records or whois log has UNKs (tokens that let users of BERT know that this version of the model does not understand a specific character or sub-word). The final form of our input vector of each domain is a concatenated vector of the domain's whois log BERT embeddings vector and its domain collection BERT embeddings vector. In the end we had 11,777 negative instances and 9,425 positive instances.

*(3.7 How the data is utilized)* Our proposed system has two components: a private,local model that does not share its gradients with others and a federated,global model that shares its gradients. The private model only uses a filter list of domains solely set/created by the user and no domains from any third-party list or project are added to it - the goal of this model is block content that the user does not want to see for subjective reasons(e.g not wanting to see Microsoft owned or associated domains). The global model's objective is to block generally agreed upon targets(e.g malware,ads,trackers). This model utilizes a base list of domains that everyone has but a user can always add lists and domains to their local version of this model(that shares its gradients with others) as well. So for this reason, we only train and test this global model(and its associated experiments) in this paper as that is the only model we can accurately represent due to the relatively objective nature of its ground truth. Furthermore, some users in our system may choose not to utilize the private content blocker and thus the global model only system can be seen as the core/base system that everyone will have access to. For the reasons above, only the global/federated model and its associated experiments get trained with the data we cleaned and processed in the steps articulated within this section.

*(3.8 Data Analysis)* We went through the filter lists of the positive class(i.e malicious content blocking filter lists) and we set aside all the projects that were hosted on GitHub. We got around 52 different projects to analyze. We gained two important insights into the overall behavior of these projects. One being that it takes around a month for the filter lists to get updated. However, larger projects tend to skew this analysis as shown by the highlighted EasyList and No-Coin projects. We can assume from this behavior that larger projects act as anchors for ad blockers whereas the smaller niche lists that get rarely updated can possibly can cover some of the more rarer deficiencies found in the larger lists(e.g a smaller list could help block more rare types of malware domains not listed in a larger project).

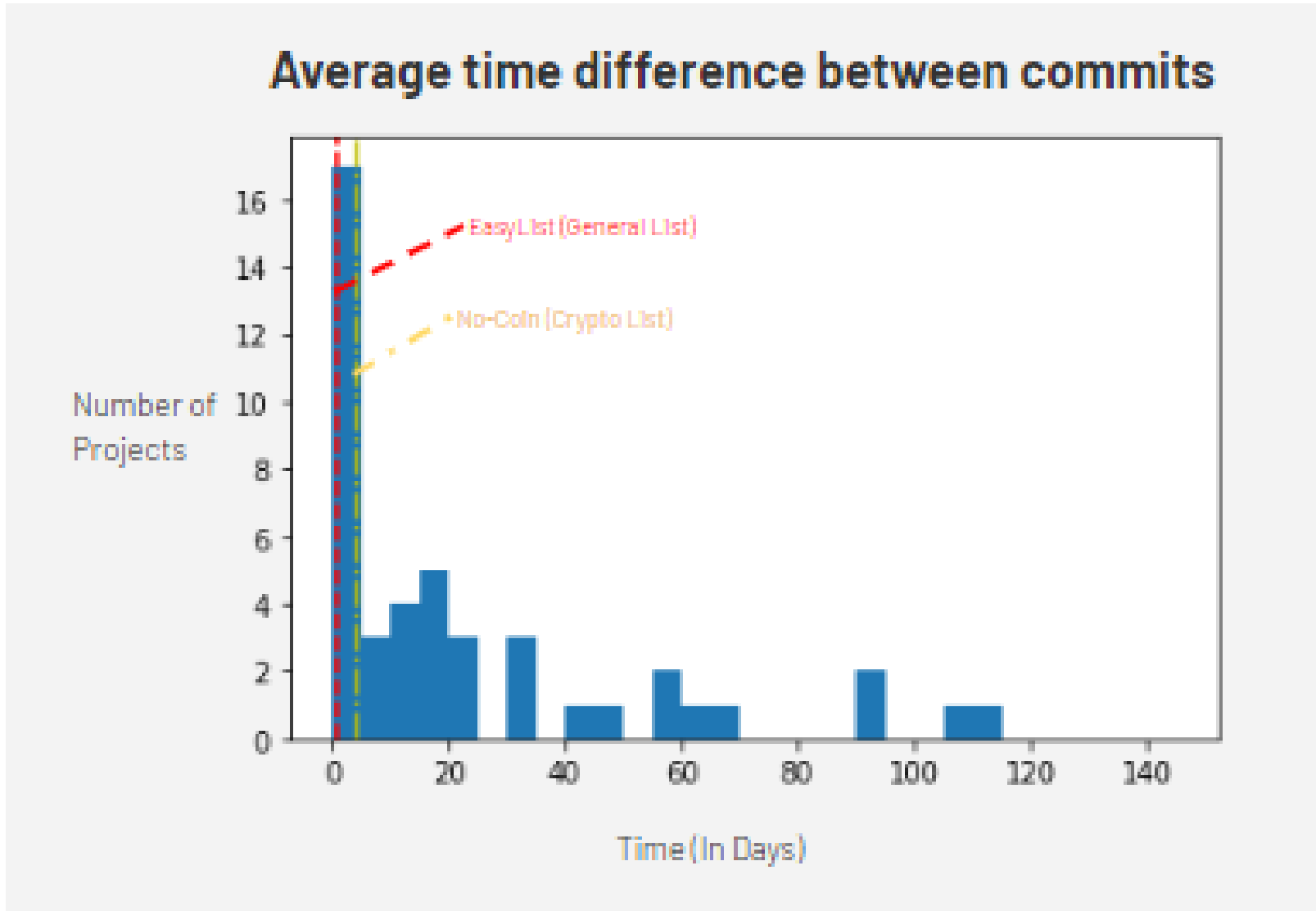

Figure 1: The figure above shows the average time difference between commits by highlighting the average commit delay/time-difference for individual projects.

Another observation we made is that most projects have very few maintainers. This means a very small group of people actually support and work on these projects. This can lead to issues of projects flatlining in terms of updates when the few users lose interest in them. Furthermore, this also possibly exacerbates the bias issue of these lists. Very few people actually decide what goes into these lists that millions of people use on a daily basis. Moreover, it would be very difficult to get volunteers from regions with less active users of content blocking technology,which would help counteract any possible regional bias in the lists.

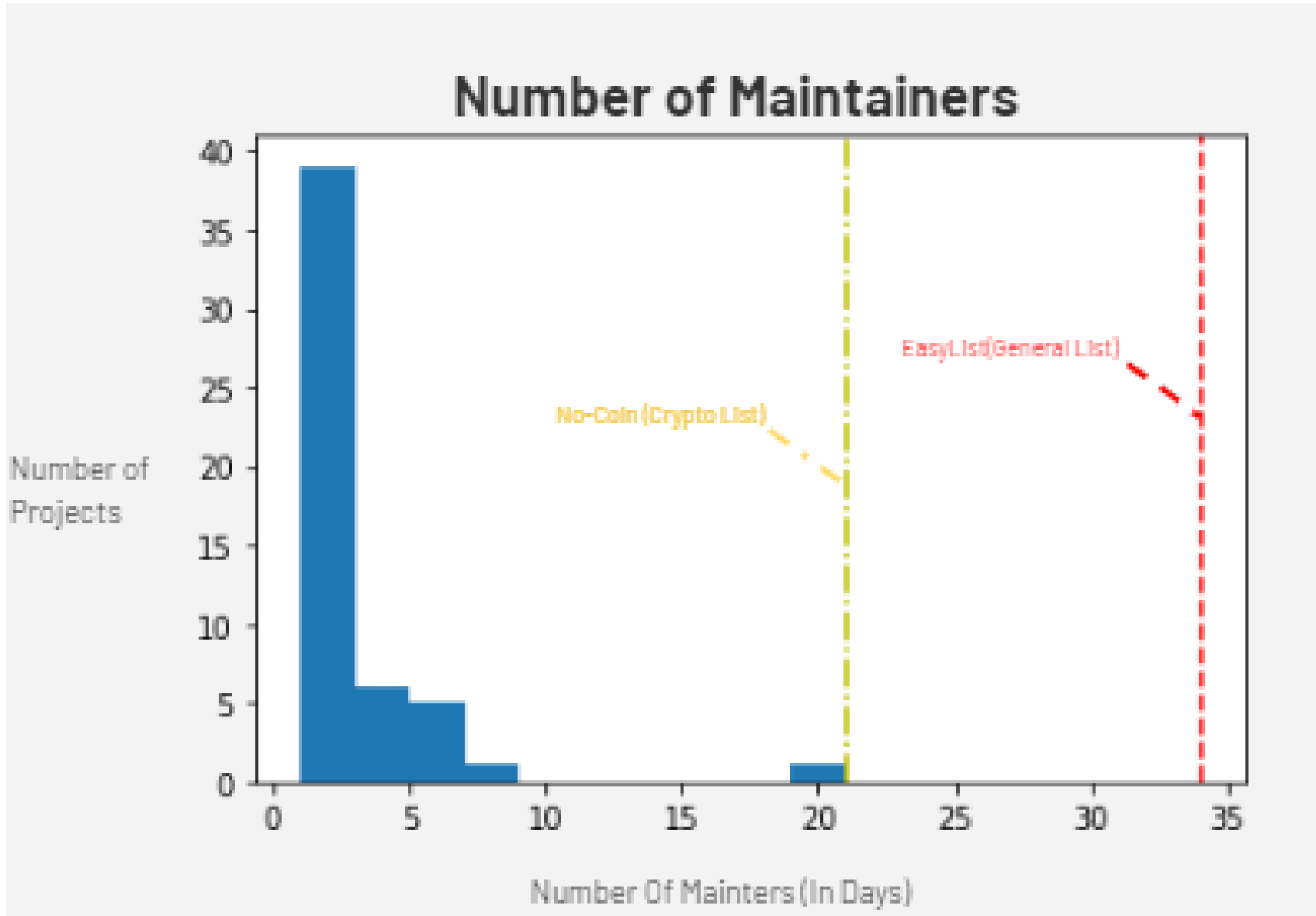

Figure 2: The figure above shows the trend displayed between open-source filter list projects with regards to the number of maintainers involved with the project.

*Overview of Data used for this analysis can be found in the Appendix (A.3)*

**4. Method: System Architecture & Design:**

*(4.1 System level objectives)* Our system-level/technical objectives are as follows. We want to create a system that can accurately inhibit malicious domains at the DNS level throughout a network for all devices within this network. The system should automatically learn from its own locally downloaded and/or updated filter lists and also from the distributed patterns seen by other users in the system. We also want the system to be relatively customizable by end users with regards to what domains it blocks on said users network. Finally, the required system should also attempt to secure itself from potential malicious attacks that could lower the effectiveness of the system and stop privacy leakages that could expose a user's private data or preferences they have entered or use to maintain or run this system.

*(4.2 Outline Of Architecture And Implementation)* Our system will act like a DNS proxy service. In other words it will take DNS queries from a local network and forward requests to a public DNS server. However, it will only forward requests it thinks are non-malicious(i.e non ad,tracker,malware,crypto miner domains) and block all malicious requests. There are three phases/components for deciding how to classify something as malicious in our system. Each local instance of our system will have a base filter list of the categories we mentioned earlier. If a domain query matches with a domain in this base list, we block said request. If a domain is not found within the base list, we then look to our two neural network based classifiers. We cache and collect the domain's associated DNS records(CNAME and AAAA) and its whois log as input for the two neural networks. If either one of the models classifies the domain as malicious, we block the domain. One neural network is trained purely on domains supplied by the user who sets up the local DNS proxy service our system is built around. This model tries to block any content users do not want to see on a network. The other system is trained on the base list we mentioned earlier and gets updated by a central service that aggregates the training gradients of different users in the system to create the updates to this model. The role of this model is to block generally malicious domains from being accessed on a network.

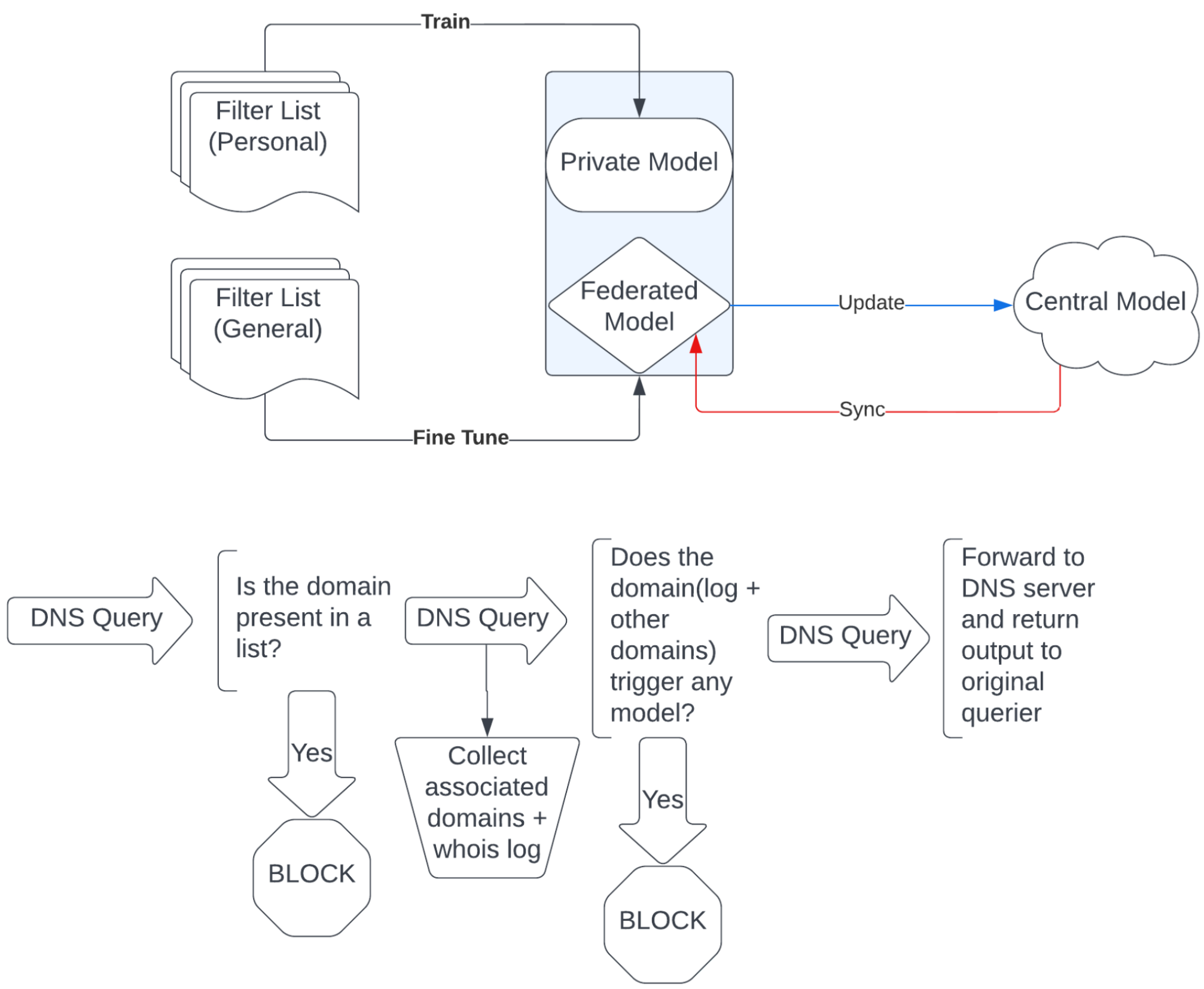

Figure 3: Overview of how an individual instance/user of the system's models are updated and how a query is handled by the system.

*(4.3 Outline of the neural networks used)* The private, general content blocker neural network and the federated,malicious content blocker neural network will share the same network architecture. We found the following architecture to be most suitable by performing randomized hyperparameter optimization on a task utilizing the data we cleaned and processed(see the "Data - Collection & Processing & Patterns" section for more details). The task being to predict whether a domain is malicious or not using the BERT embeddings of the domains alternative domains and whois log. Now in terms of architecture, both models have 5 dense MLP layers with 1536 , 416, 32 ,1 neuron(s) respectively. The two hidden layers use the ReLU and SeLU activation function respectively and the final layer uses a Sigmoid activation function. We also use binary cross entropy as our loss function and we use stochastic gradient descent with a learning rate of 0.01. The inputs to these two models is the same, the concatenated BERT embeddings of a domain whois log and all its associated domains. The output is a probability estimate on whether it is a malicious/blocked domain(i.e a label of 0) or it is a non-malicious/unblocked domain(i.e a label of 1).

*(4.4 How the Federated Model Works)* Each user gets a local-copy of the federated model that gets trained on a base filter list of domains oriented towards generally accepted malicious categories. However, a user can also add their own domains matching these categories to this list. This hits our first objective of being customizable. We characterize our federated learning system's training in terms of "rounds". The definition of which can be changed at each instantiation of the system. We define a round of training as a single step of updates across the system. At a given round of training, a random subset of users are chosen. These users are then told to go through 5 epochs of training to update a central model ,with the same architecture as their local models, using their own base lists that they may or may not have added their own domains to, The gradients from this training are aggregated on a central server and are applied to this central model. The users that are not selected at this time/round are allowed to finetune their local copies of the central model on their own base lists. After a couple of rounds(in our case 30), the local copies of the central,federated model are synced with the central model stored on the aggregation server.

*(4.5 How Our Objectives Are Met With the Federated Model)* The sharing of gradients in the manner described above also ensures patterns in unique domains across users are shared across the system whilst not physically sharing filter list rules. The fine-tuning between these sync events allows tail users that share very little in common with other users in terms of domains in their base list to still take advantage of their unique domains[20]. These two ideas ensure that we can see a boost in terms of accuracy on unseen and new domains not covered by the filter lists of singular users - covering another one of our objectives. The sharing of gradients over the system also makes it distributed in nature and the automated rounds of training ensure the system keeps up to date with the trends exhibited by the filter lists of users within the system. This also fulfills another set of our objectives. Moreover, cryptographic techniques like secure aggregation can be used to share gradients over a network without leaking said gradients[21]. Finally, we also ensure that malicious end users do not attack the accuracy of our system by flooding large gradient updates to disturb the central model by keeping a running average of gradients. If a malicious agent tries to give an anomalous gradient update(i.e larger or smaller than 2 standard deviations of the current average) we reject said gradient. Moreover, we recommend that teams trying to implement our system also ensure that a trusted third party is used to validate(but not log or tether to users to a fingerprint) new users joining the system. This will stop malicious agents from flooding the system with their own bots to ruin the central model. With these mechanisms we hope to ensure your privacy and security preserving system objectives.

*(4.6 The Private Content Blocker Model)* Each user also receives an optional content blocker model that gets trained on a list of domains that comes purely from the user. The gradient updates do not get shared and the list of domains do not get shared either. This allows users to block domains (and thereby content on them) without having their preferences leaked. This additional model makes the system more flexible on the user end. We mainly experiment and evaluate a system that purely relies on the federated model described above though. This is due to it being very difficult to model the content preferences of users on a simulated network and the optional nature of this model.

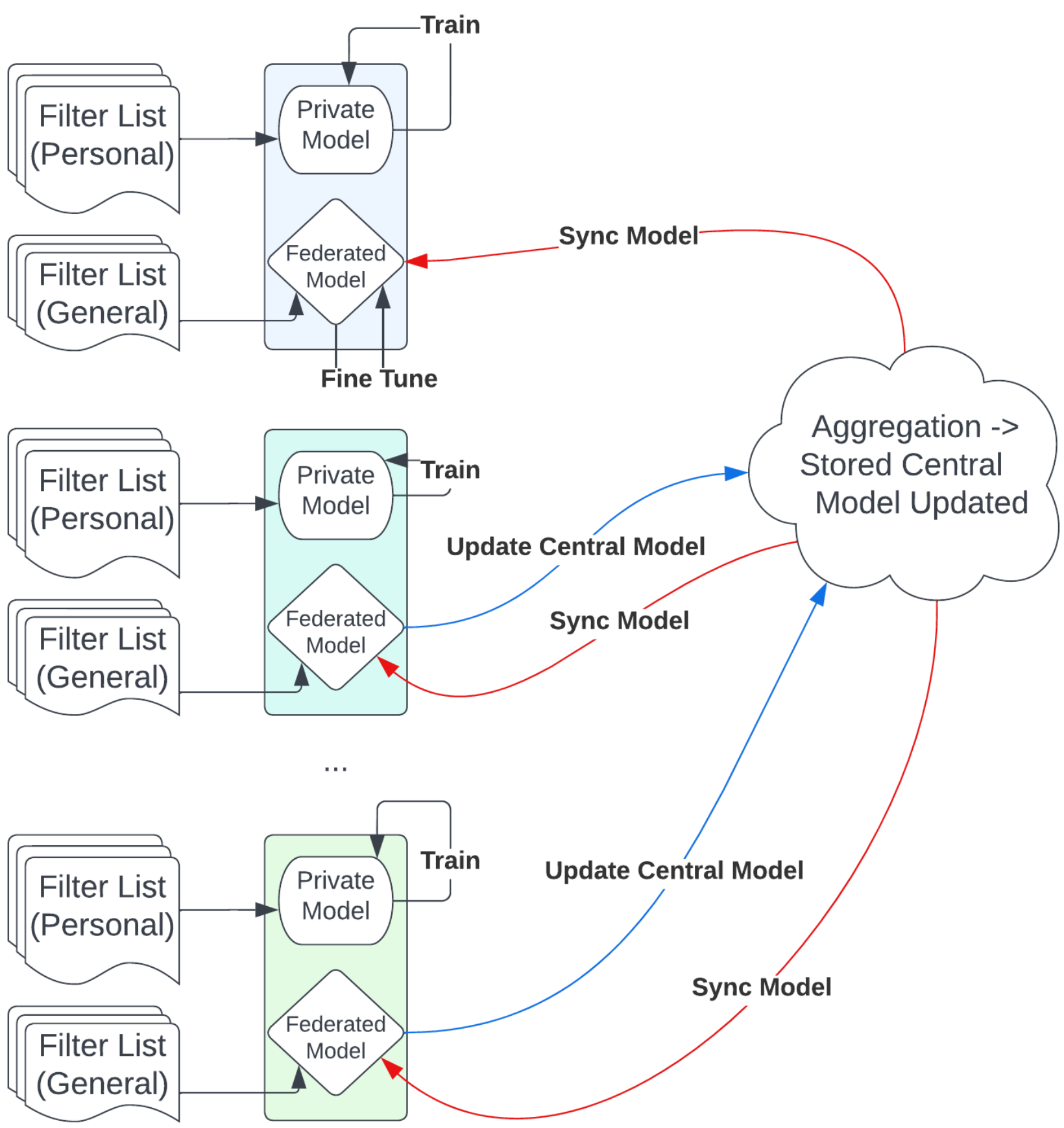

Figure 4: An overview of how the various models in the system update themselves.

## 5. Evaluation

*(5.1 Overview of tests)* We took a tiered approach in evaluating the core components and ideas of our proposed system. We first compared a baseline machine learning algorithm ,that used hand-picked features, to our BERT-fed neural network approach. The results showed that our neural network approach did indeed have comparable results to the core approach undertaken by prior work. We then experimented with federated learning and showed that a system that utilizes federated learning will outperform the pure neural network model we introduced in the previous experiment. Finally, once we had shown that our system formulation had verifiable gains, we analyzed a couple of core federated learning hyperparameters to explore their effect on the system and test the assumptions we had of the federated learning component of our system.

*(5.2 Neural Network VS RandomForest)* As alluded to in our related work section, there has been a lot of work done on blocking ads and/or trackers with machine learning techniques that use hand-picked features. The core component of our system is a neural network that takes in BERT embeddings of the whois log and associated domains - a complete departure from the previous work we have discussed. So to see if we could achieve comparable results, we compared our system's neural network architecture to a RandomForest decision tree with the features listed in the table below.

| **Random Forest Features** |
| --- |
| • Admin country,Tech Country, Registrant country on whois log(if error value set to -1) <br> • Days since domain creation on whois log (if error value set to -1) <br> • Days since expiry on whois log(if error set to 0) <br> • [Number of english words in all associated domains] / [Total number of ngrams in all associated domains] (if error value set to -1) <br> • Client and Server EPP STATUS codes present in whois log (one-hot encoded) |

Table 3: The features used to train the random forest algorithm on our dataset.

We provided our neural network (with the same architecture described in the method section) and the RandomForest the same training data described in the data section. The task was also the same: give a binary label on whether or not to block the given domain using the provided input features. Cross-validation was also performed on the RandomForest whereas a fixed validation set was used during the training of the neural network. The best models of each algorithm were picked via randomized search. On a test set ,that was set aside from the main training data, the neural network outperformed the RandomForest. One a secondary test set that was constructed using two filter lists not used in the base training data set, the neural network outperformed the RandomForest again.

| Algorithm | Accuracy | ROC Value | F1 Score |
|---|---|---|---|
| Our Neural Network | 80% | 83% | 80% |
| RandomForest | 78% | 80% | 77% |

Table 4: Metrics(rounded to the nearest whole percent) of the given algorithms on the base dataset’s test set.

| Algorithm | Accuracy | ROC Value |
|---|---|---|
| Our Neural Network | 93% | 93% |
| RandomForest | 89% | 89% |

Table 5: Metrics(rounded to the nearest whole percent) of the given algorithms on the combination of two new filter lists not used in training. One was a block list(joe wein threat feed)

and the other a allow list(baluptons-cdn-allow list). There were a total of 55 domains in the list(and their associated features). There were 26 block list domains and 29 allow list domains.

*(5.3 Federated Model Versus Non-Federated Model - Setup)* Now that we have shown the comparable performance of a neural network based approach to the task, we further investigate the usefulness of the federated component of our system. The main question to be asked here is if there is any point in adding a federated learning system on top of the neural network to further improve its performance. The federated system we created and experimented with had a central model that would be updated by randomly selected participants in the network. After a set amount of training rounds, the users' local models (that get fine-tuned on their local data when they are not selected for central model updates) get resynced with the central model - ensuring everyone has both a chance to get the latest updates from the central model and the ability to fine tune to their own models. We also created an equivalent set of non-federated models with the same architecture that do not share their gradients and purely train on their own personal data. Both systems received training data from the same training data set from the previous experiment(neural network vs RandomForest). Each user of both systems gets a "base list" of training data instances they all share in common. We created sub-experiments/(system instances) that changed the number of simulated clients in the system and the number of unique domains added on top of the base list per client. Furthermore, we also experimented with how fast the federated system converges when given a new set of data. By examining and combining the results of all these sub-experiments and configurations of the system, we were able to analyze the performance of the federated and unfederated systems.

*(5.4 Federated Model Versus Non-Federated Model - Performance Results)* We created 9 sets of experiments/configurations(that we repeated thrice) where we changed the number of clients in the system and number of unique domains per user to see how they impacted the federated and non-federated systems' performance. We test the performance of the non-federated and federated system in each of these experiments and then analyze the results as a whole from these experiments(i.e we look at all the results from all the experiments to compare system performance). Testing the performance of the system this way allows us to analyze performance of the system in various configurations and limits possible biased setups that unfairly give an edge to the federated or non-federated system. The exact results of these experiments can be found in the appendix(A.1). In each experiment the training dataset was partitioned to the hyperparameters alluded to earlier and the resulting models were given the same test set of 55 domains(which was the secondary test set of the baseline vs neural network experiment). Again, in each experiment each user was given a federated model that they fine tuned for a couple of rounds before it was reset and a non-federated model trained purely on their local dataset that was assigned to them. Since we have multiple instances of two models in each experiment (one set for each user in the system), we considered a model type(i.e federated or non-federated) to outperform another in a given experiment if the average accuracy of the models of one type was higher than that of the other in that instance of the system or experiment. The average accuracy was fixed to one standard deviation below the mean(i.e if average accuracy for a set of non-federated models was 50 and the standard deviation was 2, we used an accuracy of 48 for comparison). Using this as a basis of comparison, we found that in around 74% of experiments/configurations, the federated model that was fine-tuned outperformed the equivalent private models. The fine-tuned model we used for this analysis were federated models the users had just before the final resync with the main model. Using the same scheme for comparison, we also found that the final set of non-synced federated models(i.e federated model right before the final resync) only beat the central model in around 19% of experiments. However, the gains of a non-fine tuned central model begin to wane when we consider the fine tuned models beat the central model in around 78% of experiments/configurations when the test set was switched to each user's local dataset. This is vitally important for tail users with very unique domains in their filter lists as they are at least guaranteed some localized performance for taking part in the system[20]. The results therefore show that our configuration of having a fine-tuned federated model gives users in our system the best of both worlds(performance and localized adaptation for tail users).

*(5.5 FL Hyperparameter Impacts On Performance - Client Size and Unique Domains)* It is also important to note how the parameters we changed in each experiment/configuration affected the performance of the system. The first observation to take note of is the interplay between the number of clients in the system versus the number of unique each client has. The figure below shows that starting from the 10-50 client size, we see a relatively loose pattern: the fewer unique domains we have the better the overall performance of the system. This pattern becomes clear and linear in larger client sizes. The lower client sizes also seem to achieve better performance,especially when the unique domain size is very small. However, we see that for the lower client size experiments a "bucket curve" pattern emerges: where having more unique domains is better after a certain cutoff point and having less is even better after a certain cutoff point. With our fixed 150 round training, the more unique data there is in the system, the more competing gradients we will see that will be saying different things and thus it will be harder to imbue the "collective knowledge" within such a system when there is too much flux in the system. A possible solution is to increase the training time. Increasing the training time for higher client sizes, will result in more stable performance improvement whereas doing so for lower client sizes might result in mixed performance. Therefore an ideal system would be: a lot of clients, not too much variation between clients, and infinite training time. This a very realistic assumption to make as each of these points can be easily expressed in a real world implementation of our system.

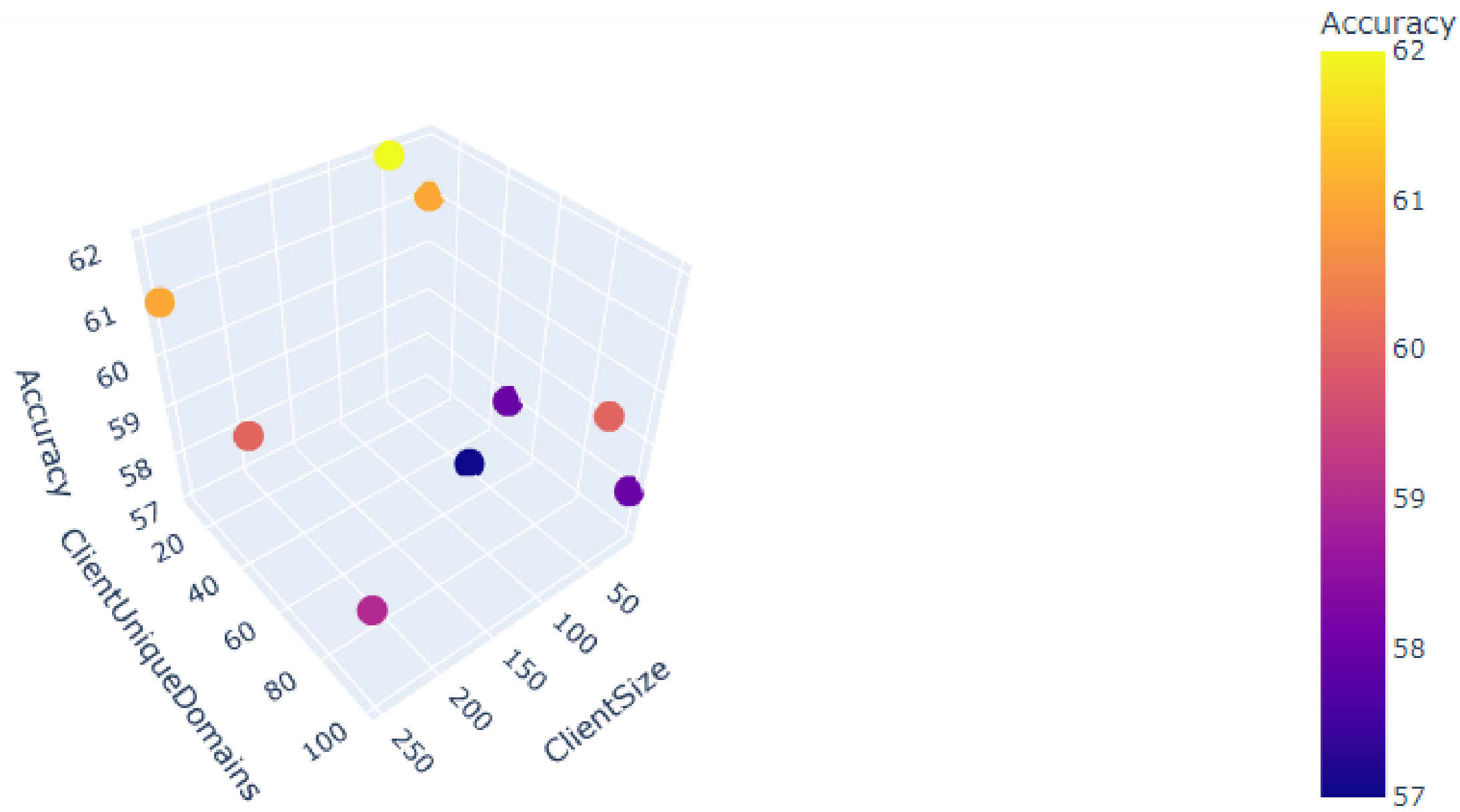

Figure 5. This diagram shows the average lowest bound accuracy(average mean accuracy of each experiment minus one standard deviation from said mean) of each of the 9 experiments with the number of clients(ClientSize) and unique domains per client(ClientUniqueDomains) of each experiment being highlighted as x and y axes.

*(5.6 FL Hyperparameter Impacts On Performance - Convergence)* Since we expect this system to have an infinite runtime with updates rounds and fine tuning, we also expect occasional updates to occur on the ground truth (base filter lists of each user). So it is vital to see how different configurations of the system react to such updates. For each of the 9 experiments/configurations mentioned earlier, we also had a secondary set of experiments(3 for each of the 9 experiments) where we changed the number of added domains after the system had been fully trained and examined how the system reacted to them. After removing any configurations that did not have a clear loss improvement(as we want to pick out the best configurations), the top experiments (i.e ones that had the best loss improvement calculated on the new domains) are those with ones with less domains to add. We found that the fewer the number of clients and fewer number of base unique domains, the better the loss improvement.

So a system with few new updates and  little uniqueness from user to user and many update rounds is most suitable. All in all, we recommend small updates in a realistic system in order to help improvements trickle out into the system.

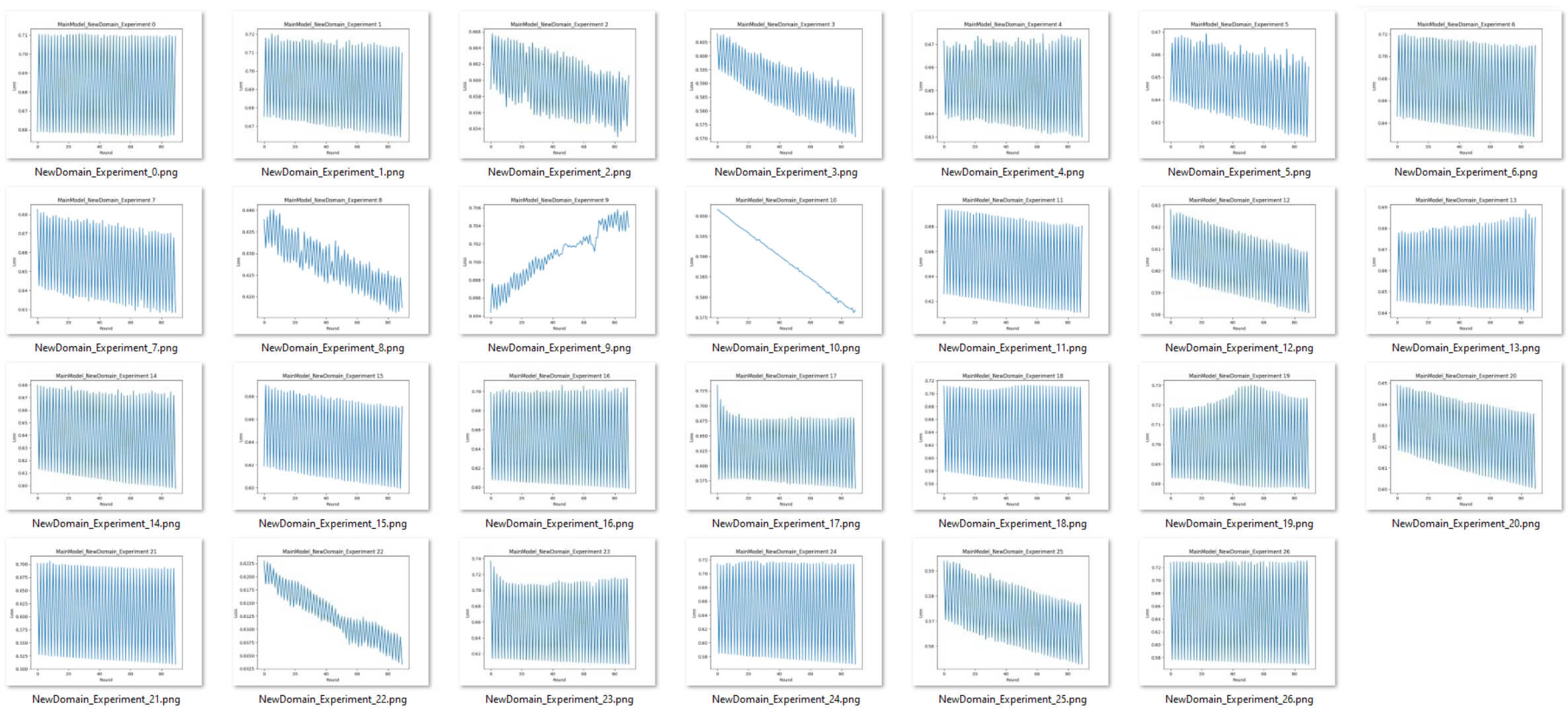

Figure 6. The loss graphs of the federated system once the new domains were randomly added(the number of each is based on the experiment's parameters). The loss is calculated based on the given new domains that were added to the trained federated models (and by extension central model).

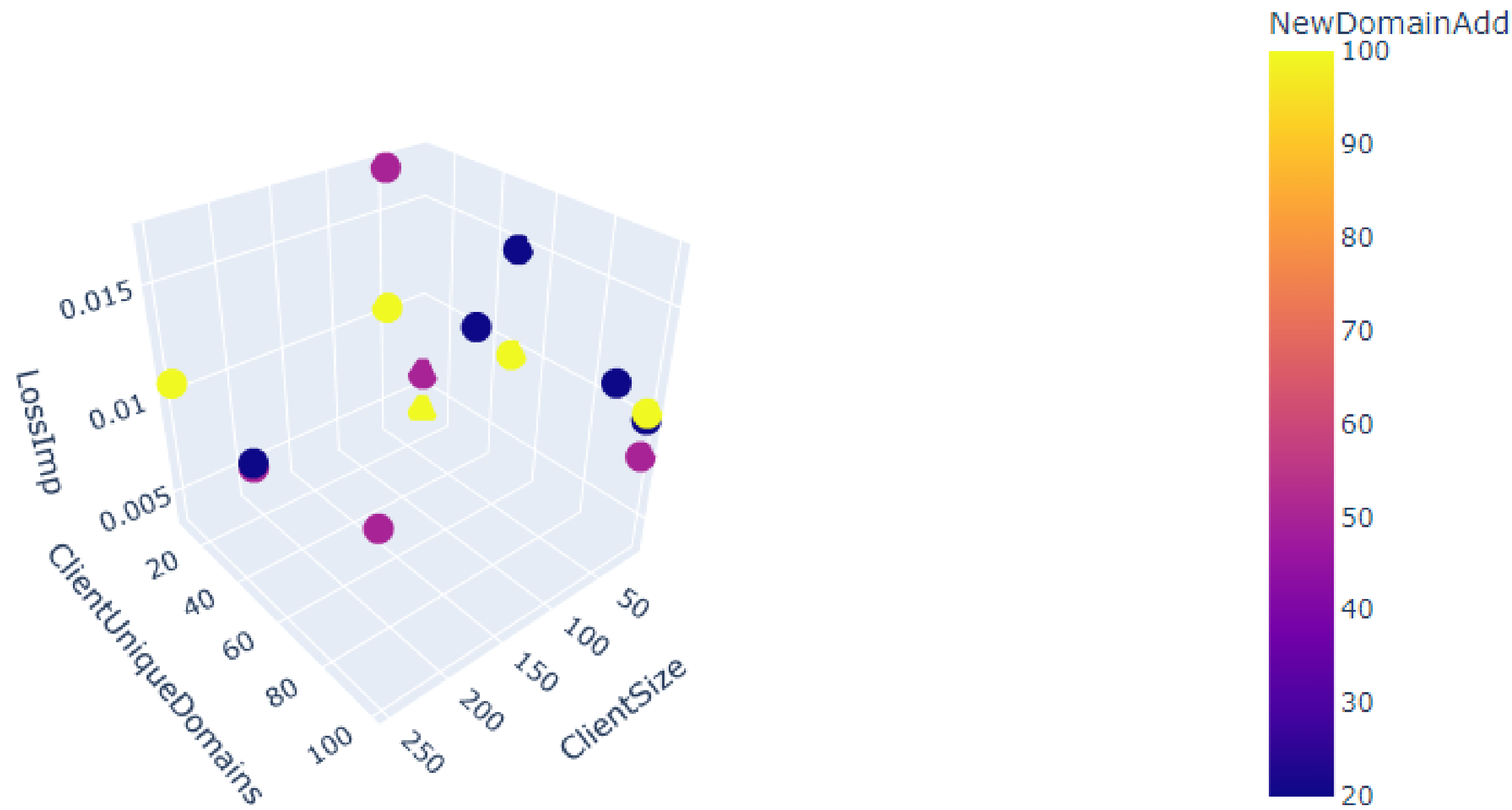

Figure 7. This diagram shows the loss improvement of each of the models that were selected. The x and y axes show the number of clients in the system and unique domains per client. The coloring represents the number of domains added after the initial models were trained. Tabular data for the above can be found in the appendix (A.2)

*(5.7 Key takeaways)* The main takeaway is that the system we have formulated has appreciable potential. We have shown that it outperforms the ideas of previous work and that the added complexity of introducing federated learning into the system stands to improve the overall system. We have also noted some recommendations on how to properly manage the different aspects of a system in an applied setting.

**6. Discussion & Future Work:**

*(6.1 Explainability)* Though we use a transformer to obtain embeddings that we then feed into our neural networks, we do not have any direct way of deducing why the model works the way it does since the transformer is not an active component of our neural network architecture. It is more of a data transformation technique in our pipeline. When it comes to adoption use this might be an issue as filter lists provide direct insight on what is being blocked. A solution that might augment our accuracy as well, is to use a transformer possibly in conjunction with a neural network to classify our textual inputs [22]. These will give us  the ability to employ interperability techniques geared towards transformers and will, in turn, allow us to highlight portions of the whois log and domains that trigger a model to classify it as either malicious or non-malicious.

*(6.2 Alternative Architectures and Embeddings)* As highlighted earlier, we primarily use a basic multi-layer perceptron as our primary model and BERT as our primary driver for text representation: this gives future researchers ample room to experiment with other formulations and architectures. To possibly boost performance, architectures like recurrent neural networks (e.g LSTMs, RNNs with GRU units) can be used with more traditional word embeddings like GLoVE vectors. More complicated language models (e.g GPT-3) can also be used or possibly fine-tuned to acquire a better vectorized representation of our textual input.There is no limit here when it comes to architectural innovation for this task due to how simple the target is and how much possible data we have to discriminate between domains.

*(6.3 Counter-blocking threats)* Counter-blocking of ad blockers is a real threat to the validity of the proposed system. Back in 2016, Nithyanand et al. first analyzed the presence of anti-ad blocker measures on websites and found that around 7% of the top 5,000 websites (according to the Alexa Top-5K list) employed anti-ad blocking measures[23]. We can only assume that this number has grown alongside the increased adoption of ad blockers across the web. If a website can detect a user is actively using our system to block certain aspects of their page, they can force the user to disable our system to access their website. Though dealing with such threats is more of an implementation level issue and beyond the scope of this paper, it is still important to note the possible drawbacks this system can face on the open internet.

*(6.4 Centralization,Consensus, and the Filter Bubble)* Since there are no competing servers that offer aggregation services in a centralized system(i.e only one central server and model) for the users taking part in the federated learning process, whatever the majority of users decide on becomes the “ground truth” for the system. This is due the fact that aggregated gradients that are used to update the central model come directly from a random assortment of users. If a large portion of the users share a similar set of gradients ,due to their similar blocking preferences, with the system during the training phase of the federated model then the entire system moves in the majority’s favor as they will be selected more often due to their large presence within the system.Though the fine-tuning of the system offers more unique users of the system some baseline guarantee of performance, it does not ensure that their idea of malicious/non-malicious domains will be shared across other users in the system. For these users that have completely different views on what should be blocked or unblocked, the resync with the central model that occurs every so often, only drags the performance of their local systems down. Essentially, such a system will create its own “filter bubble”. Users who have the same philosophy as the majority on what should be blocked will tend to stick around longer whereas those who have completely divergent views will tend to leave more frequently. This cycle only increases the divergence between users which ends in a large portion of users possibly leaving if the decisions the majority takes on what should be blocked is very divisive. The most direct and simple solution would be to create different central servers for each varying geographical region - assuming that users residing in the same region share the same opinion.

In another proposal, private individuals might also create and advertise their own central model services. This gives users the ability to choose servers that are more inline with their own ideals.

*(6.5 Effect of ad blocking on websites)* According to Shiller et al (2017), using ad blocker(and by association: derivative technology like we are suggesting in this paper) indirectly harms the livelihood of the websites that take advantage of them to garner revenue[24]. By using such technology that blocks the presence of ads on a webpage, the website does not receive payment for showing said ads to that specific user. This results in lower short profits that result in reduced investment into the further development of the website: lowering quality[24]. This reduction in quality further pushes away users from using this website and ultimately results in a loss of potential long-term revenue required to stay up on the internet[24]. This spells disaster for many privately run websites that depend on these ads to pay for hosting costs. This has a negative effect on the web economy.

*(6.6 Is it ethical to block ads?)* As mentioned previously, there is an economic cost associated with blocking ads. However, not blocking said ads opens users to a whole host of privacy related issues. These issues mainly stem from the use of trackers that track user behavior in order to suggest the most pertinent ad that a user will most likely click. According to Iqbal et al.(2017), these trackers allow private firms to extract sensitive information (e.g medical conditions and financial state) from users [25]. Furthermore, governments and other institutions can take advantage of this private extracted data in order to perform unethical surveillance on unknowing citizens[25]. So a moral quandary appears: hurt small and large businesses that depend on these ads or risk leaking private information. The burden of choice,therefore, on whether or not to utilize the system we are presenting within this paper falls on the potential user. They need to perform a cost-benefit analysis based on their own set of ethics before using any such system.

## 7.Conclusion & Acknowledgements:

*(7.1 Performance)* Our goal was to create a novel system that would be an effective supplement for the current filter list based content blocking ecosystem. We feel we have achieved this objective by showing that our isolated model has comparable performance to a simulated baseline and showing that using federated learning further augmented the performance of this isolated model.

*(7.2 Foundation For Future Work)* We would also like to add that this federated system and the ML model used as components for this system can be further explored and improved. Therefore, this paper is also a potential foundation for future work looking into using machine learning to improve domain blocking systems in a privacy preserving and decentralized manner.

*(7.3 Explainability)* Outside further performance improvements, the main goal of future work should be to look into adding explainability into the system to facilitate possible real-world adoption of this system. The most promising angle of this challenge is to look into using transformers and their self-attention layers to highlight what aspects of the input text are triggering the model.

*(Acknowledgement)* I would like to thank Professor Z. Shafiq for his continued and unwavering support during the course of this project. I would also like to thank Professor. S. Rafatirad and Professor. M. Bishop for their support and guidance as well. A special thanks to VirusTotal for the use of their whois data I used for training and WHOISXML for the use of their web categorization and whois data that I used for exploratory data analysis. Finally, I would like to thank my parents and my friend Taimur (and anyone I have not pointed out by name) for helping me on this journey.

**Citations:**

**Appendix:**

- Table 6: (A.1) Section Reference: **Evaluation**

| Experiment Number | Experiment Configuration | FL Central Model Final Acc on Test Set | Private Models average (Acc,Stand Dev) on Test set | Fine Tuned Avg (Acc,Stand Dev) on Test set | Fine-tuned FL models avg (Acc,Stand Dev) on own local data | Central FL model avg (Acc,Stand Dev) on own local data |
|---|---|---|---|---|---|---|
| 0 | [10, 10, 20] | 58.18 | (69.45, 6.63) | (60.18, 2.63) | (59.22, 0.86) | (58.14, 0.09) |
| 1 | [10, 10, 50] | 61.82 | (64.73, 6.37) | (69.45, 4.6) | (59.43, 1.44) | (60.76, 0.12) |
| 2 | [10, 10, 100] | 69.09 | (58.36, 3.48) | (67.09, 5.71) | (60.26, 1.32) | (58.92, 0.16) |
| 3 | [10, 50, 20] | 78.18 | (61.64, | (66.73, | (63.61, | (60.65, |
| | | | 5.84) | 7.43) | 1.53) | 0.26) |
| 4 | [10, 50, 50] | 81.82 | (61.45, 7.94) | (72.55, 10.89) | (60.81, 1.31) | (59.46, 0.28) |
| 5 | [10, 50, 100] | 58.18 | (65.82, 9.06) | (66.0, 11.75) | (61.7, 1.17) | (62.22, 0.3) |
| 6 | [10, 100, 20] | 70.91 | (64.36, 7.53) | (64.0, 7.85) | (60.99, 1.43) | (58.4, 0.56) |
| 7 | [10, 100, 50] | 58.18 | (69.09, 9.35) | (64.55, 6.01) | (60.39, 1.24) | (60.94, 0.4) |
| 8 | [10, 100, 100] | 74.55 | (67.27, 8.35) | (67.09, 7.49) | (61.01, 0.64) | (60.35, 0.45) |
| 9 | [50, 10, 20] | 52.73 | (65.78, 8.28) | (65.27, 10.86) | (61.15, 1.95) | (63.28, 0.12) |
| 10 | [50, 10, 50] | 74.55 | (65.93, 7.18) | (76.25, 0.77) | (69.15, 0.65) | (67.1, 0.13) |
| 11 | [50, 10, 100] | 56.36 | (64.69, 8.59) | (62.07, 5.7) | (60.6, 2.01) | (56.6, 0.14) |
| 12 | [50, 50, 20] | 70.91 | (64.95, 6.95) | (62.91, 8.09) | (61.27, 0.9) | (59.63, 0.39) |
| 13 | [50, 50, 50] | 61.82 | (65.45, 7.5) | (65.02, 5.55) | (62.45, 1.46) | (62.93, 0.37) |
| 14 | [50, 50, 100] | 63.64 | (65.2, 9.05) | (62.22, 4.64) | (60.46, 1.69) | (58.29, 0.36) |
| 15 | [50, 100, 20] | 67.27 | (63.67, 6.7) | (64.62, 6.8) | (61.05, 1.57) | (57.93, 0.45) |
| 16 | [50, 100, 50] | 70.91 | (65.45, 8.27) | (67.38, 5.15) | (61.26, 1.83) | (58.07, 0.43) |
| 17 | [50, 100, 100] | 74.55 | (65.16, 7.06) | (66.84, 8.28) | (63.33, 1.33) | (61.04, 0.51) |
| 18 | [250, 10, 20] | 76.36 | (65.04, 7.62) | (68.76, 7.47) | (61.06, 0.99) | (59.49, 0.15) |
| 19 | [250, 10, 50] | 72.73 | (65.28, 7.79) | (65.28, 7.35) | (61.86, 1.71) | (59.21, 0.15) |
| 20 | [250, 10, 100] | 72.73 | (65.13, 7.56) | (69.28, 5.94) | (61.01, 1.28) | (59.41, 0.15) |
| 21 | [250, 50, 20] | 67.27 | (64.88, 7.98) | (64.39, 6.37) | (62.34, 2.13) | (59.28, 0.33) |
| 22 | [250, 50, 50] | 72.73 | (65.45, 7.94) | (69.29, 8.99) | (63.61, 0.95) | (61.08, 0.32) |
| 23 | [250, 50, 100] | 74.55 | (65.15, 7.59) | (69.33, 8.56) | (61.32, 1.64) | (57.16, 0.33) |
| 24 | [250, 100, 20] | 76.36 | (65.35, 8.06) | (69.0, 8.13) | (61.46, 1.51) | (59.54, 0.44) |
| 25 | [250, 100, 50] | 60 | (65.74, 7.54) | (68.78, 9.83) | (62.04, 1.39) | (61.71, 0.46) |
| 26 | [250, 100, 100] | 67.27 | (65.05, 7.85) | (63.11, 6.71) | (60.78, 1.12) | (58.59, 0.45) |

Table 6: The accuracy (and standard deviation) for the different federated system experiments organized by model. Experiment configuration organized as follows: [Number of Clients,Unique Domains Per Client,Number Of Domains Added To Check Convergence]

- Table 7: (A.2) Section Reference: **Evaluation**

| Experiment Number | Number Of Clients | Unique Domain Per Client | New Domains Added | Loss Improvement |
|---|---|---|---|---|
| 2 | 10 | 10 | 100 | 0.0036 |
| 1 | 10 | 10 | 50 | 0.0058 |
| 7 | 10 | 100 | 50 | 0.0082 |
| 5 | 10 | 50 | 100 | 0.0096 |
| 22 | 250 | 50 | 50 | 0.0098 |
| 6 | 10 | 100 | 20 | 0.01 |
| 21 | 250 | 50 | 20 | 0.01 |
| 8 | 10 | 100 | 100 | 0.0103 |
| 11 | 50 | 10 | 100 | 0.0103 |
| 20 | 250 | 10 | 100 | 0.0106 |
| 25 | 250 | 100 | 50 | 0.0117 |
| 12 | 50 | 50 | 20 | 0.0118 |
| 15 | 50 | 100 | 20 | 0.0126 |
| 3 | 10 | 50 | 20 | 0.0147 |
| 10 | 50 | 10 | 50 | 0.0171 |

Table 7: The table shows the loss improvements of the experiment where we add new domains once the models have been trained to see how they converge. The experiment number corresponds to the table A.1.

- Table 8: (A.3) Section Reference: Data

| ID | Title | Description | Type | User | Project Name | Working Branch | FilePath | Average Commit Time | MaintainerNum | MostCommonTypeOfChange | NumCommits |
|---|---|---|---|---|---|---|---|---|---|---|---|
| 0 | Yuki's uBlock Japanese filters - Mobile | Add this if you use Yuki's uBlock Japanese filters with uBlock Origin on Firefox for Mobile. | MOBILE | Yuki2718 | adblock | master | japanese/jp-mob.txt | 0.796969 | 2 | MODIFY | 97 |
| 2 | gmbkslist (Hosts) | Blocks Japanese regional mobile advertisements | MOBILE | DandelionSprout | adfilt | master | AncientLibrary/gmbkslist (Hosts).txt | -1 | 1 | ADD | 1 |
| | | and trackers. | | | | | | | | | |
| 7 | Anti-PopAds | Blocks shady, annoying pop-under ads from the infamous PopAds ad network. | IOT | Yhonay | antipopads | master | popads.txt | 0.140408 | 4 | MODIFY | 11857 |
| 9 | ABP Japanese 3rd Party SNS Filters | Blocks Japanese regional social network advertisements and trackers. | IOT | k2jp | abp-japanese-filters | master | abpjf_3rd_party_sns.txt | 67.548644 | 1 | MODIFY | 22 |
| 12 | Minimal Hosts Blocker | A minimal adblock | MOBILE | arcetera | Minimal-Hosts-Blocker | master | etc/MinimalHostsBlocke | 7.922361 | 3 | MODIFY | 31 |
| | | er for iOS. | | | | | r/minimalhosts | | | | |
| 13 | Adawaylist JP | Hosts for Adaway. Block mainly on advertisements for mobile. | MOBILE | multiverse2011 | adawaylist-jp | master | hosts | 90.28502 | 1 | MODIFY | 8 |
| 14 | All-in-One Customized Adblock List | A comprehensive, all-in-one adblock list for thorough blocking of trackers, popup ads, ads, unwanted cookies, | IOT | hl2guide | All-in-One-Customized-Adblock-List | master | aio.txt | 0.956952 | 1 | MODIFY | 96 |
| | | fake news, cookie warning messages, unwanted comment sections, crypto-coin mining, YouTube clutter and social network hassles. | | | | | | | | | |
| 15 | PiHole Blocklist SmartTV | None | IOT | Perflyst | PiHoleBlocklist | master | SmartTV.txt | 16.38762 | 21 | MODIFY | 76 |
| 16 | PiHole Blocklist Android Tracking | None | IOT | Perflyst | PiHoleBlocklist | master | android-tracking.txt | 55.656209 | 6 | MODIFY | 18 |
| 17 | Badd Boyz Hosts | This list of hosts is compiled from server logs on my own servers and forms the basis of the bad referrers domain lists for The Nginx Ultimate Bad Bot Blocker at https://github.com/mitchellkrogza/nginx-ultimate-bad-bot-blocker | IOT | mitchellkrogza | Badd-Boyz-Hosts | master | hosts | 1.720956 | 6 | MODIFY | 893 |
| | | and the Apache Ultimate Bad Bot Blocker at https://github.com/mitchellkrogza/apache-ultimate-bad-bot-blocker | | | | | | | | | |
| 19 | Goodbye Ads | Goodbye Ads is designed for Unix-like systems (such as Android), gets a list of domains that serve ads, | MOBILE | jerryn70 | GoodbyeAds | master | Hosts/GoodbyeAds.txt | 18.187391 | 2 | ADD | 65 |
| | | tracking scripts and malware from multiple reputable sources and creates a hosts file that prevents your system from connecting to them. | | | | | | | | | |
| 21 | pihole-ytadblock | YouTube Ad-Block-List for PiHole by Henning VanRäumle | IOT | Henning VanRaumle | pihole-ytadblock | master | ytadblock.txt | 0.761523 | 1 | MODIFY | 82 |
| 23 | Adguard Mobile Ads (hosts) | None | MOBILE | r-a-y | mobile-hosts | master | AdguardMobileAds.txt | 10.207265 | 1 | MODIFY | 155 |
| 24 | Adguard Mobile Tracking and Spyware (hosts) | None | MOBILE | r-a-y | mobile-hosts | master | AdguardMobileSpyware.txt | 13.551155 | 1 | MODIFY | 117 |
| 26 | Spotify AdBlocking for pihole | None | IOT | w13d | adblockListABP-PiHole | master | Spotify.txt | 4.084722 | 1 | ADD | 2 |
| 28 | Facebook Zero Hosts Block | This aim to block non-formal hosts that serve all Facebook contents and resources from alternative "Free | MOBILE | kowith337 | PersonalFilterListCollection | master | hosts/hosts_facebook0.txt | 24.012658 | 1 | MODIFY | 45 |
| | | Basics" servers that it happen when you're using on mobile data over the carrier that collaborate with Facebook to have THAT service! This list will follow the update after routine check results of dead-hosts. | | | | | | | | | |
| 29 | Mat1th DNS add block (Domains) | None | IOT | deathbybandaid | piholeparser | master | Subscribable-Lists/Parsedblocklists/Mat1th-DNS-add-block.txt | -1 | 1 | ADD | 1 |
| 33 | CitizenX VIL Hosts Mobile | Mobile ad/tracker list based on AdGuard's mobile ad filter. | MOBILE | CitizenX VIL | Hosts | master | mobile domains.txt | 55.003472 | 1 | MODIFY | 10 |
| 34 | tankmohit's AllUnifiedHosts | Unified hosts combined from wally3k pihole lists and possibly without false positives | IOT | tankmohit | Unified Hosts | master | hosts.all | 45.384722 | 2 | RENAME | 2 |
| 35 | DNS Zone block list Generator or Dnsmasq | This project generates dnsmasq, bind and unbound zone files to be used in DNS based AD Blockers. | IOT | oznu | dns-zone-blocklist | master | dnsmasq/dnsmasq.block list | 3.689965 | 1 | MODIFY | 454 |
| 36 | SNAFU List | A general ad & tracking domain blocklist for pihole. | IOT | Rooney McNibNug | pihole-stuff | master | SNAFU.txt | 2.446188 | 8 | MODIFY | 393 |
| 37 | DataMaster Android AdBlock Hosts | Android AdBlock Hosts file for /etc/hosts | MOBILE | DataMaster-2501 | DataMaster-Android-AdBlock-Hosts | master | hosts | 105.339331 | 1 | MODIFY | 12 |
| 38 | Mobile-Ad-Hosts | This ad blocker list aims to block mobile ads which includes in-app ads. | MOBILE | biroloter | Mobile-Ad-Hosts | master | hosts | 94.067361 | 1 | MODIFY | 13 |
| 39 | hosts-jp Ads | Categorized hosts files for DNS based content blocking. This is meant to be used as a regional component of a more comprehensive hosts | IOT | tiuxo | hosts | master | ads | 42.798997 | 2 | MODIFY | 19 |
| | | list, such as Steven Black's hosts. | | | | | | | | | |
| 40 | PiHole Lists - Quad9 | None | IOT | XionKzn | PiHole-Lists | master | PiHole/Archive/Quad9.txt | -1 | 1 | RENAME | 1 |
| 41 | PiHole Lists - Yahoo Ad Servers | None | IOT | XionKzn | PiHole-Lists | master | PiHole/Archive/Yahoo_Ad_Servers.txt | -1 | 1 | RENAME | 1 |
| 42 | PiHole Lists - Spyware | None | IOT | XionKzn | PiHole-Lists | master | PiHole/PiHole_HOSTS_Spyware_HOSTS.txt | 147.195139 | 2 | MODIFY | 3 |
| 43 | PiHole Lists - Cerber Ransomware | None | IOT | XionKzn | PiHole-Lists | master | PiHole/Archive/Cerber_Ransomware.txt | -1 | 1 | RENAME | 1 |
| 44 | PiHole Lists - Blocklist | None | IOT | XionKzn | PiHole-Lists | master | PiHole/Blocklist_HOSTS.txt | 15.054094 | 2 | MODIFY | 20 |
| 45 | iOS Tracker Blocklist | My personal blocklist of iOS tracking, telemetry, and advertising domains. | MOBILE | jakejarvis | ios-trackers | master | blocklist.txt | 32.930469 | 1 | MODIFY | 17 |
| 46 | Dandelion Sprout's AdGuard Home Compilation List | AdGuard Home is one of the tools for the future, or at least it is for those who know how to set up and use a Raspberry Pi. | IOT | DandelionSprout | adfilt | master | AdGuard Home Compilation List/AdGuardHomeCompilationList.txt | 16.711762 | 1 | MODIFY | 49 |
| | | However, its current internal structure discards virtually every−*usingblock rule*(*e.g. EasyList*)*,whichmakesEasyListandsimilarlists virtually uselessin AdGuard Home*,*despitehow manyoftheentrieswouldbe*≥90values, here's my intermissional list to work | | | | | | | | | |
| | | around it. | | | | | | | | | |
| 47 | Dandelion Sprout's Anti-Malware List (for AdGuard Home) | Most anti-malware lists are pretty big and can cover a 5- or 6-digit amount of specific domains. But my list hereby claims to remove more than 25% of all known malware sites with just | IOT | DandelionSprout | adfilt | master | Alternate versions Anti-Malware List/AntiMalwareAdGuardHome.txt | 5.716025 | 1 | MODIFY | 145 |
| | | a 2-digit amount of entries. This is mostly done by blocking top-level domains that have become devastatingly abused by spammers, usually because they allowed for free and uncontrolled domain registrations. There's | | | | | | | | | |
| | | also additional categories that cover unusual malware and phishing domains that very few other lists seem to cover. | | | | | | | | | |
| 48 | Dandelion Sprout's Nordic Filters for Tidier Websites (for AdGuard Home) | Removes ads, affiliation ads, and empty boxes on Norwegian, Danish, Icelandic, Sami | IOT | DandelionSprout | adfilt | master | NorwegianExperimental List alternate versions/NordicFiltersAdGuardHome.txt | 5.273381 | 1 | MODIFY | 146 |
| | | and Danish territorial sites to produce a cleaner browser experience. Meant to be used on top of general filters. uBO users should use the regular version instead. | | | | | | | | | |
| 49 | ADP Mobile Filter | None | MOBILE | T4Tea | ADPMobileFilter | master | ADPMobileFilter.txt | 10.31313 | 1 | MODIFY | 55 |
| 50 | AdditionalFiltersCN | Additional Filters for browser extension based adblockers like uBlock Origin, mainly for Chinese mainland internet service, respectively for privacy, advertisement and interruption elements. Including some obvious trackers | IOT | Crystal-RainSlide | AdditionalFiltersCN | master | CN.txt | 23.099599 | 1 | MODIFY | 27 |
| | | which could be found&confirmed by novices in network development easily, and should been listed&blocked YEARS ago. | | | | | | | | | |
| 52 | mhhakim Pihole Blocklist | Custom blocklist to include in the main list. | IOT | mhhakim | pihole-blocklist | master | custom-blocklist.txt | 112.53125 | 2 | MODIFY | 5 |
| 53 | Samsung No Snooping | List below is all domains I've | IOT | UnbendableStraw | samsungsnooping | master | README.md | 4.48998 | 1 | MODIFY | 8 |
| | | captured that Samsung Smart TVs use to do anything other than watch YouTtube, Netflix, and Google Play. Block all of them for total privacy and no ads. Note the comments. | | | | | | | | | |
| 54 | xlimit91 Block List for Pi-Hole | Custom block list annoying ads, | IOT | xlimit91 | xlimit91-block-list | master | blocklist.txt | 34.582812 | 3 | MODIFY | 25 |
| | | trackers, scam sites etc. for Pi-hole (DNS Blocking) | | | | | | | | | |
| 55 | Th3M3 Blocklists - Malware | My custom blocklists for pihole | IOT | Th3M3 | blocklists | master | malware.list | 12.103538 | 1 | MODIFY | 22 |
| 56 | Th3M3 Blocklists - Tracking & Ads | My custom blocklists for pihole | IOT | Th3M3 | blocklists | master | tracking&ads.list | 34.061111 | 1 | MODIFY | 14 |
| 57 | Game Console Adblock List | Much like there's now lists for AdGuard Home and Pi-hole to block ads on smart-T | IOT | DandelionSprout | adfilt | master | GameConsoleAdblockList.txt | 24.68529 | 2 | MODIFY | 23 |
| | | Vs, here's an attempt from me at doing the same for videogame consoles with AdGuard Home. Enjoy. | | | | | | | | | |
| 61 | dnswarden Adblocking (Domains) | Blocklist mainly used for dns servers at dnswarden | IOT | dnswarden | blocklist | master | blocklist-formats/hostnames | 0.760646 | 3 | MODIFY | 116 |
| 62 | anti-AD (Domains) | anti-AD is currently the highest-hitcount | IOT | privacy-protection-tools | anti-AD | master | anti-ad-domains.txt | 1.274459 | 4 | MODIFY | 498 |
| | | -filtering ad list in Chinese, which achieves accurate ad blocking and privacy protection. Now supports AdGuard Home, dnsmasq, Surge, Pi-Hole and other excellent network tools. | | | | | | | | | |
| 66 | ADgk Mobile Advertis | None | MOBILE | banben dalao | ADgk | master | ADgk.txt | 2.333449 | 3 | MODIFY | 199 |
| | ing Rules - adgk手机去广告规则 | | | | | | | | | | |
| 67 | PiHole Blocklist SmartTV - Amazon Fire TV | None | IOT | Perflyst | PiHoleBlocklist | master | AmazonFireTV.txt | 349.675694 | 2 | MODIFY | 3 |
| 68 | Combined Privacy Block Lists HOSTS (IPv4 + IPv6) | This is a comprehensive hosts file (IPv4 + IPv6) which blocks known ad, exploit, malware, and tracking servers. It is pulled from | MOBILE | bongochong | CombinedPrivacyBlockLists | master | newhosts-final-Dual.hosts | 2.437594 | 2 | MODIFY | 391 |
| | | MVPS, PGL Yoyo, Malware Domain List, Energiz ed and EasyList , along with some supplem entary entries for increase d protecti on against telemetr y, and addition s for mobile platform s. It is then merged, sorted | | | | | | | | | |
| | | and dedupe d. This list is compati ble with all operatin g systems that make use of a hosts file (obvious ly this includes Window s, OS X, GNU/Li nux and more). It works great with mobile VPN ad-block ing | | | | | | | | | |
| | | solutions too. | | | | | | | | | |
| 71 | Host-List for iOS ad blockers (Domains) | For use with AdGuard Pro | MOBILE | BlackJack8 | iOSAdblockList | master | Hosts.txt | 18.903571 | 5 | MODIFY | 64 |
| 72 | blackbook | blackbook is a historical (black)list of malicious domains created as part of the periodic automated heuristic check (i.e. WHOIS, HTTP, etc.) of newly | IOT | stamparm | blackbook | master | blackbook.txt | 1.139991 | 1 | MODIFY | 676 |
| | | reported entries from public lists of malicious URLs (currently CyberCrime, URLhaus, ScumBots, Benkow and VirusTracker). Main goal is listing those that are/were malware dedicated (e.g. C&C) - thus, | | | | | | | | | |
| | | excluding compromised sites. It is supposed to be used for detection of malware beaconing infected clients by inspection of associated DNS traffic, with significant reduce of false-positives. | | | | | | | | | |
| 74 | AdGuard DNS Filter - Additional rules | One of AdGuard DNS Filter's many source files. This one contains entries for domains not covered by any of its other source lists. | IOT | Adguard Team | AdGuardSDNS Filter | master | Filters/rules.txt | 64.12116 | 6 | MODIFY | 18 |
| 75 | AdBlock List (Domains) | Self-Customized AdBlock List Focusing on DNS hijacking and | IOT | Licolnlee | AdBlock List | master | AdList | 0.441667 | 1 | MODIFY | 3 |
| | | AdBlock. | | | | | | | | | |
| 76 | AdBlock List Lite (Domains) | Self-Customized AdBlock List Focusing on DNS hijacking and AdBlock. | IOT | LicoInle | AdBlock List | master | AdListLite | 0.884028 | 1 | RENAME | 2 |
| 80 | Adblock by Jm (Hosts) | I created this project as a way to optimize adware protection of my router TPLINK 1043 with Openwrt | IOT | jmhenrique | adblock | master | etc/adblock_hosts | 2.47715 | 5 | MODIFY | 777 |
| | | . It's an excellent router, but has very little available memory (8MB) and a median processor (400MHz). I noticed that the articles and tutorials on ad-blocking does not take into account the optimization of hosts | | | | | | | | | |
| | | and domains. (…) But these lists are variations of websites that create random subdomains, interfering with the blocking efficiency. Only one domain, 302br.net has +17,000 registered subdomains in lists. So I first | | | | | | | | | |
| | | tried to treat (sub)domains within the router, (…) and impacted the performance of the navigation here at home, (…). This takes ~2Mb on the router. After grouping, my list has about 27,000 hosts. (…) | | | | | | | | | |

Table 8: The table shows the data used for analyzing the average time between commits and the number of average number of maintainters. Note that repos with only one commit were excluded from the average time between commits graph.